\documentstyle[prl,aps,epsf]{revtex}
\draft
\begin{document}
\twocolumn[\hsize\textwidth\columnwidth\hsize\csname @twocolumnfalse\endcsname

\title{Glassy Ratchets For Collectively Interacting Particles}
\author{C. Reichhardt, C.J. Olson Reichhardt, and  M.B. Hastings} 
\address{ 
Center for Nonlinear Studies and Theoretical Physics Division, 
Los Alamos National Laboratory, Los Alamos, NM 87545}

\date{\today}
\maketitle
\begin{abstract}
We show that ratchet effects can occur in a glassy media 
of interacting particles where there is no quenched substrate. 
We consider 
simulations of a disordered binary assembly of colloids in 
which only one species 
responds to a drive. 
We apply an asymmetric 
ac drive that would produce no net dc drift of an
isolated overdamped particle.
When interacting particles are present,
the asymmetric ac drive produces a ratchet effect.
A simple model captures many of the 
results from simulations, including flux reversals
as a function of density and temperature. 
\end{abstract}
\vspace{-0.1in}
\vspace{-0.3in}

\vskip2pc]
\narrowtext
In ratcheting systems, a dc current or flux arises
upon application of 
a periodic drive, such as when the underlying
potential substrate is flashed, or, for a fixed
substrate, when alternating forces are applied \cite{Reimann1}. 
Ratchet effects, which can be deterministic or
stochastic in nature, 
have been studied in a wide variety 
of systems, including biological 
motors \cite{Astumian2}, 
flux motion in superconductors \cite{Janko3},
granular transport \cite{Granular4}, 
and colloidal transport. 
In many of these systems,
an underlying asymmetric 
potential acts to break the left-right spatial symmetry. 
Alternative ratchet systems that have recently been investigated
\cite{Hastings5} have {\it symmetric} substrates, and rely on
ac drives to provide symmetry breaking.  In each of these
cases, the particles move through some form of fixed substrate.

In this work we study a ratchet effect that arises
due to the {\it collective} interactions between 
driven and non-driven particles in a system that 
does not contain a fixed substrate.  
The particular system we consider is a binary mixture 
of interacting repulsive particles which form a 
disordered or glassy media. 
These particles, referred to as the {\it passive} species, 
interact only with other particles, and do not respond
to an applied ac drive.  We introduce a small number of a separate
species of particles, termed the {\it active} species, 
which respond both to the other particles and
to the ac drive. 
Using numerical simulations we demonstrate 
that a ratchet 
effect can be achieved for the active particles
in this system
when the coupling between the active and passive
particles is strong enough
and a symmetry breaking
is introduced by means of the ac drive.  
This is in contrast to a recent theoretical proposal of an asymmetric
drive ratchet which transports undriven superconducting
vortices \cite{PRatchet}.
In addition, for specific ac drives, we find
current reversals as a function of density and temperature. 
We also show that the magnitude of the ratchet effect goes through a maximum 
as a function of the number of active particles,
due to the collective string-like motions of the
particles.  

We explicitly demonstrate a
ratchet effect for  mixtures of colloids.    
Our system is similar to recent 
studies of binary colloidal models driven
with ac fields where one 
species of the colloids moves in the 
direction of the applied drive while the other species moves in the opposite 
direction \cite{Lowen6Lowen27}.
These studies have found interesting 
collective effects in which
the moving particles organize into lanes.
In extensive studies of driven diffusive models \cite{Zia8}, 
where the two species of oppositely moving
particles are placed on a lattice,
a rich variety of nonequilibrium phases were observed. 
Continuum models, with different portions of particles moving
in opposite directions, have also been 
studied in the context of pedestrian flows \cite{Helbing9}. 
A remarkable phenomena termed ``freezing by heating,'' 
where the system can jam at higher temperatures,
has recently been
observed in these types of systems \cite{Vicsek10}.  
Since  
individual colloids or groups of colloids can be manipulated easily with 
optical traps \cite{Grier11},
colloidal systems are ideal for experimentally realizing a
system in which only a fraction of the particles
respond to the applied drive. 
The active species could be created using charged or magnetic
colloids, while the passive particles would be charge-neutral or
non-magnetic.
For example, in recent experiments,
magnetic colloids were driven with ac \cite{Weeks112} 
and dc drives \cite{Weeks213} through a glassy assembly of other non-magnetic 
colloids. 
It should also be possible to drive a fraction
of the colloids using optical or magnetic 
traps. 

We consider a binary two-dimensional (2D) assembly of colloids using 
the same simulation method as in our previous works 
\cite{Reichhardt14,Periodic15}.  
The overdamped equation of motion for colloid $i$ is
\begin{equation}
\frac{d {\bf r}_{i}}{dt} = {\bf f}_{ij} + 
{\bf f}_{AC} + {\bf f}_{T}\ ,
\end{equation}
where ${\bf f}_{ij} = -\sum_{j \neq i}^{N}\nabla_i V(r_{ij})$ 
is the interaction force from the other colloids, 
${\bf f}_{AC}$ is the force due to an applied driving field,
and ${\bf f}_{T}$ is a thermal force from Langevin kicks.   
The colloids interact via a Yukawa or screened Coulomb
interaction potential, 
$V(r_{ij}) = (q_{i}q_{j}/|{\bf r}_{i} - {\bf r}_{j}|)
\exp(-\kappa|{\bf r}_{i} - {\bf r}_{j}|)$.
 Here $q_{i}$ is the charge on 
colloid $i$, $1/\kappa$ is the screening length, and ${\bf r}_{i (j)}$ 
is the position of
particle $i (j)$.
The system size is measured in units of the lattice constant 
$a$, which is set to $a=a_0$ for most of this work,
giving a particle density $\rho = 1/a_{0}^{2}$.
We take the screening length $\kappa = 3/a_{0}$. Our system 
is periodic in both the $x$ and $y$ directions. 
The total number of particles is $N = N_{A} + N_{B} + N_{D}$,
with $N_{A}$ colloids of passive species A and charge $q_{A}$,
and $N_{B}$ colloids of passive species B and charge $q_{B}$. 
We add $N_{D}$ active colloids 

\begin{figure}
\center{
\epsfxsize=3.5in
\epsfbox{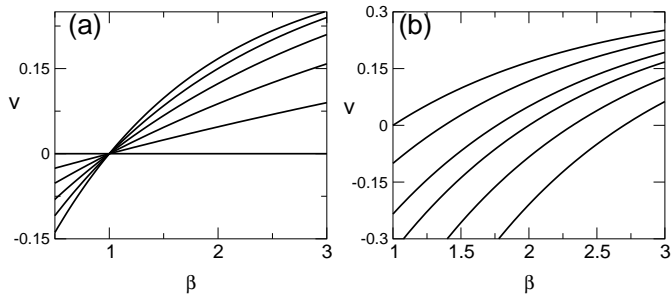}}
\caption{
(a) Velocity $v$ vs power-law exponent $\beta$ from Eq.~2 for $f_{1} = 1.0$.
Ratcheting in the positive direction ($v>0$) occurs for
$\beta > 1$. From top left to bottom left, 
$f_{2}/f_{1} = 1.0$ (flat line), 0.9, 0.8, 0.7, 0.6,
and $0.5$. (b) $v$ vs $\beta$ for fixed $f_{1}/f_{2} = 0.5$ and, 
from bottom to top, $\tau_{2}/\tau_{1} = 3.25$, 2.5, 2, 1.5, 1.25, and $1$.  
}
\end{figure}

\noindent
with charge $q_{D}$ that respond
to an applied ac drive. 
For the results presented here, we fix $q_{A}/q_{B} = 1/2$.
We have 
also considered other ratios, 
including the crystalline case when $q_{A}/q_{B} = 1$. 
To initialize the system, we start at a high temperature 
and anneal either to $T = 0$ or to a fixed $T>0$. The binary
particles of species A and B form a disordered assembly, 
with the active colloids scattered randomly throughout.
After annealing, we apply a two-step ac drive
during a single cycle of period $\tau$. 
The first stage is a {\it fast} drive where a 
constant force of $f_{1}$ is applied for a time $\tau_{1}$
in the positive $x$-direction.  The second stage is a
{\it slow} drive where a force $f_{2}<f_{1}$ is applied
for a time $\tau_{2}>\tau_{1}$ in the negative $x$-direction. 
The ac drive has the properties $f_{1}\tau_{1} = f_{2}\tau_{2}$, 
with $\tau = \tau_{1} + \tau_{2}$, so that
the particle experiences zero average force during a single cycle.
The average drift velocity $v_d$
can be obtained after many cycles. We 
conduct a series of simulations for different
particle densities, system sizes, temperatures, 
$q_{D}$, ac amplitudes, frequencies, and $N_{D}$. 

We first consider a 
simple model that predicts that a ratchet  
effect can arise in such a system. 
Recently it was shown theoretically and
in simulations that, for a single colloid
driven with a dc drive through a
glassy assembly of other colloids \cite{Reichhardt14}, 
the colloid velocity follows
a power law $V \propto f^{\alpha}$ with $\alpha = 1.5$
for certain parameters. 
The power law appears in regimes where the driven colloid produces
collective disturbances of the surrounding colloids
on length scales of the order of $a$, the lattice constant. 
When the
density of the surrounding colloids is low, these collective distortions
are lost and the velocity of the driven colloid is linear 
with the drive.  Linear behavior
also occurs when the charge $q_D$ of the driven particle is
much smaller than that of the surrounding particles.
Recent experiments on driven magnetic colloids have also found
power law velocity-force relations for the driven colloid 
with exponents $\alpha = 1.5$ or higher
in systems with high density \cite{Weeks112}.  In
low density systems, the velocity is linear with drive. 

We find that for a 
{\it group} of driven particles in a glassy media, 
the velocity-force curve is also of a power law 

\begin{figure}
\center{
\epsfxsize=3.5in
\epsfbox{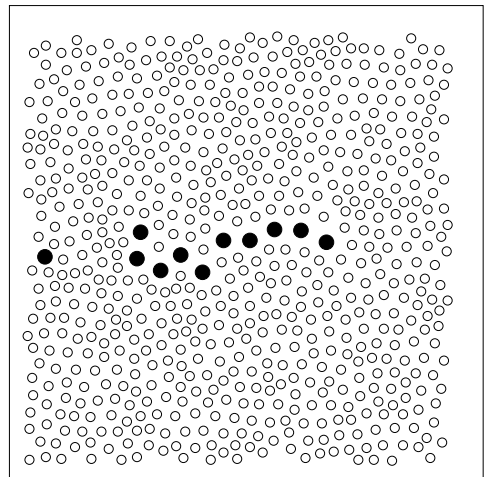}}
\caption{Simulation image. Black circles: active particles 
that respond to the ac drive.  Open circles: passive particles forming
the glassy background. Here, $N_D/N=0.016$.}
\end{figure}

\noindent
form.
In a system with a power law velocity-force 
relationship,
ratcheting can 
occur under the application of
an asymmetric ac drive.  
During a single cycle the colloids move a distance
$\delta x = f_{1}^{\beta}\tau_{1} - f_{2}^{\beta}\tau_{2}$. 
The net drift velocity is then  
$v_d = \delta x/(\tau_{1} + \tau_{2})$. Using 
the condition $f_{1}\tau_{1} = f_{2}\tau_{2}$ the drift velocity can be 
expressed as
\begin{equation}
v_d = \frac{f_{1}^{\beta}f_{2} -f_{2}^{\beta}f_{1}}{f_{2} + f_{1}} 
\end{equation}   
If $\beta = 1$, as in the case of a single overdamped particle, 
then $v_d = 0$, as expected.
For any value of $\beta > 1$, however, the 
drift velocity will be positive, $v_d>0$.  
Ratcheting can still occur if $f_{1}$ is close to $f_{2}$
as long as $\beta>1$.
This model predicts that the ratchet effect becomes
more efficient for larger values of $\beta$.
In Fig.~1(a) we plot $v_d$ vs $\beta$ from Eq. (2),
with $f_{1} = 1$ and
ratios of $f_{2}/f_{1} = 0.5$  to $1$. For 
$f_{2}/f_{1}=1$,
$v_d = 0$ for all $\beta$, as expected. 
For $f_{2}<f_{1}$ and $\beta > 1$, 
ratcheting in the positive direction ($v_d>0$) occurs. $v_d$ increases
with larger $\beta$ and larger $f_{2}/f_{1}$.
The velocity reverses for $\beta < 1$. 
This is an unphysical limit for the colloid system, 
where the lowest possible value 
is $\beta = 1$. However, some physical systems 
such as non-Newtonian fluids could have $\beta < 1$.   

Since the velocity becomes linear with the drive at
low densities or high drives, it should be possible to select
ac drive parameters that reverse the
ratchet effect and give current reversals for $\beta > 1$. 
This can be achieved by setting $\tau_{1} < \tau_{2}$,
where the inequality is not too large.  In the highly nonlinear
regime the particle will still ratchet in the positive direction; however,
upon approaching the linear regime, 
the particle will ratchet in the negative direction.
In Fig.~1(b) we show 
$v$ vs $\beta$ from Eqn. (2) 
for $f_{2}/f_{1} = 0.5$ and
increasing $\tau_{2}/\tau_1$, 
where a crossover from a positive to negative ratchet effect
occurs at $\beta_c$.  
Here $\beta_c$ increases with increasing $\tau_{2}/\tau_{1}$. 

We next consider simulation results. 
In Fig.~2 we show a real space image
of the simulated system with $N_D=11$ 

\begin{figure}
\center{
\epsfxsize=3.5in
\epsfbox{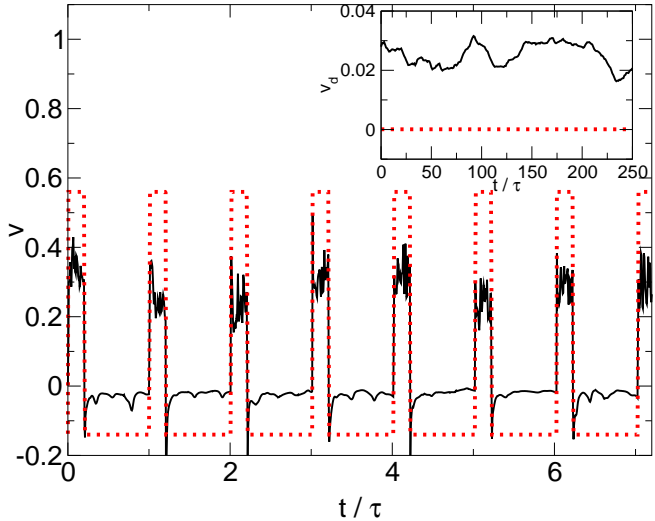}}
\caption{Particle velocity $v$ vs time $t/\tau$.
Dotted line: $v$ for
an isolated overdamped particle under an asymmetric ac drive
with $f_{1}=0.56$, $f_{2}/f_{1}=0.25$,  
and $\tau_{1}/\tau_{2} = 0.25$.
Solid line: $v$ in
a system of interacting particles. Inset: 
Drift velocity $v_d$
vs $t/\tau$ for
(dotted line) a single overdamped particle, and (solid line)
the interacting particle system. } 
\end{figure}

\noindent
active particles (black circles)
and $N_A+N_B=678$ 
passive particles (open circles) after many cycles of 
the ac drive. 
The active particles, which were originally scattered throughout
the system, have clustered into a lane, and
they also drift over time in the positive $x$-direction. 

In Fig.~3 we plot the
velocity vs time of a single overdamped colloid in the absence of
any other particles (dashed curve), showing
short large positive pulses and long small negative pulses of velocity.
Here we consider the case of $f_{2}/f_{1} = 0.25$ with $f_{1} = 0.56$. 
The average velocity of the isolated colloid
is zero in a single period. 
The solid line shows the velocity of a driven colloid
in the {\it interacting} particle system 
with $N = 370$ and $N_{D} = 11$, in a regime where the dc velocity-force
curves have a power-law form.
Here, the magnitude of the particle velocity $v$ during both the positive
and the negative portions of the ac drive cycle
is smaller than that of the single particle, and shows
considerable fluctuations.
The integrated velocity during the positive portion of the ac cycle
is on average larger in magnitude than the 
integrated velocity during the negative portion of the cycle. 
To show this more clearly, in the inset of 
Fig.~3 we plot the time-averaged drift velocity 
$v_d$ for each system over
many ac cycles. 
For the single particle case (lower dashed curve), $v_d=0$,
while for the interacting particle system, $v_d > 0$,
with
an long time average drift velocity of $<v_d> \approx 0.03$, 
indicating 
that a positive ratchet effect is occurring.   

We next consider the
dependence of the ratchet effect on various system
parameters. In Fig.~4(a) we plot
the drift velocity  $v_d$ vs the 
fraction of driven particles, $N_D/N$, 
for fixed ac drive parameters
$f_1=0.56$, $f_2/f_1=0.25$, and $\tau_1/\tau_2=0.25$,
with $a/a_0=1$. 
$v_d$ goes through a 

\begin{figure}
\center{
\epsfxsize=3.5in
\epsfbox{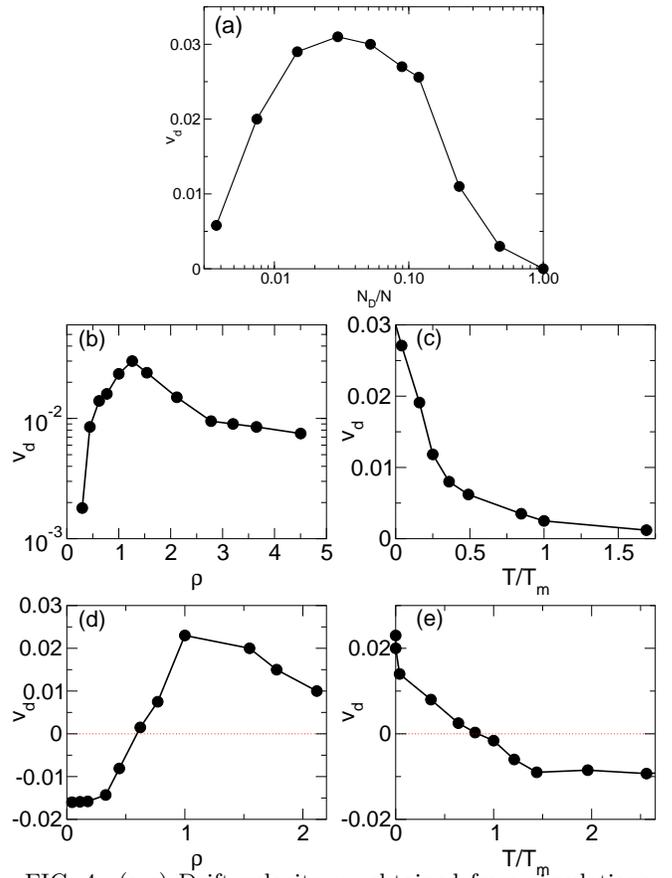}
}
\caption{
(a-c) Drift velocity $v_d$ obtained
from simulations with
$f_{1}\tau_{1} = f_{2}\tau_{2}$ and $f_{2}/f_{1} = 0.25$. (a)
$v_d$ vs the fraction of driven particles $N_{D}/N$ for 
a density of $\rho=1/a_0^2$. 
(b)
$v_d$ vs particle density $\rho$ for fixed $N_{D}/N = 0.03$. 
(c) $v_d$ vs temperature $T/T_{m}$ for $\rho=1/a_0^2$ and $N_{D}/N=0.03$.
(d-e) Ratchet reversals for a system with 
$f_{1}\tau_{1}/f_{2}\tau_{2} = 0.9$: 
(d) $v_d$ vs density $\rho$.
(e) $v_d$ vs temperature $T/T_{m}$ for $\rho=1.26/a_0^2$.
}
\end{figure}

\noindent
maximum near 
$N_{D}/N = 0.03$, and then slowly falls off for
higher $N_D/N$ before reaching $v_d=0$
just below $N_{D}/N = 1.0$. 
We have also considered different system
sizes by varying $a/a_0$.   We obtain
the same behaviors, indicating that the
ratcheting is not a finite size effect. 
It is simple to understand why 
$v_d \rightarrow 0$ in the 
limit of $N_{D}/N \rightarrow 1$, since
here all the particles move in 
unison with the ac drive, and the plasticity 
responsible for the nonlinear velocity-force response
is lost. 
For a single driven particle $N_D=1$, 
we found a dc velocity-force scaling exponent of
$\beta=1.5$. As the number of driven
particles $N_D$ increases, $\beta$
increases to $\beta=2$ near $N_{D}/N = 0.03$, and then 
decreases back toward $\beta=1$ as $N_D/N$ increases further. 
This result
is consistent with the predictions of Eq.~2, where a larger ratcheting effect
is expected for larger exponents $\beta$. 

In Fig.~4(b) we consider a system with  
fixed $N_{D}/N = 0.03$ and ac drive
parameters, and plot $v_d$ vs the system density $\rho$
measured in units of $1/a_0^2$.
For low densities $\rho < 0.5$, 
the active particles interact only weakly with the
passive particle background, and the dc velocity-force curves 
are close to linear ($\beta \approx 1$)
so little ratcheting occurs. For large densities 
$\rho > 1.3$, the ratcheting drift velocity $v_d$
decreases since it becomes more difficult for
the active particles to
pass the passive particles. 
We have also fixed the system density and varied the
charge $q_D$ of the active particles. The behavior is similar to
Fig.~4(b): the ratchet effect is lost at small $q_{D}$ when the
velocity-force curve becomes linear.
In Fig.~4(c) we show the effect of a nonzero temperature on
the ratchet effect.
We plot $v_d$ vs $T/T_{m}$, where $T_{m}$ is defined as the
temperature at which a completely ordered single-species colloid assembly 
with charge $q_{D}$ and density $\rho=1/a_0^2$
melts. For high $T$, 
the velocity increases linearly with force, 
so the ratchet effect is lost.  

Equation 2 predicts that a reverse ratchet 
effect should occur for $\beta<1$, 
as shown in Fig.~1(a), 
or for $f_{1}\tau_{1} \neq f_{2}\tau_{2}$.
The reverse ratchet effect can occur in systems where the 
media responds easily to slow moving particles but 
becomes stiffer under faster perturbations. 
For example, in granular media the grains respond in a
fluid manner to a slow moving object, while 
for a fast moving object the granular material jams and
acts like a solid.
Many polymer systems show a similar response.
In order for a ratchet
effect to occur in a substrate-free system,
there must be
a combination of the nonlinear behavior of the particle motion
and the left-right symmetry breaking
by the ac drive.
In general the ratchet effect we observe should occur for any media that
has a nonlinear viscosity-frequency response. 

In Fig.~4(d,e) we consider two cases where flux reversals occur
with $f_{1}\tau_{1}/f_{2}\tau_{2} = 0.9$. 
For the parameters chosen here, the drift velocity of an isolated particle
would be $v_d=-0.016$.
In Fig.~4(d) we plot $v_d$ vs density $\rho$
for $N_D/N=0.03$.
For low density $\rho < 0.3$, the velocity-force response curve
is linear ($\beta=1$) and the particle moves
in the negative direction as expected for this ac drive. 
As the density increases,
the velocity-force response curve becomes nonlinear,
and $v_d$ crosses over to a {\it positive}
value $v_d>0$ as the ratcheting effect reaches
a maximum near $\rho=1/a_0^2$. 
The drift velocity drops at high density just as in 
Fig.~4(b). 
In Fig.~4(e) we show 
$V$ vs $T/T_m$
for the same system with a fixed density of $\rho=1.26/a_0^2$. 
For low temperatures,
the velocity-force response curve 
is nonlinear, and a ratchet effect with positive velocity occurs.
As the temperature increases, the drift velocity $v_d$ drops and
crosses back to a negative value as the system 
enters the linear response regime.  $v_d$ saturates
to $v_d\approx -0.01$, slightly smaller in magnitude than 
the value expected for a single isolated particle.
We have also considered the effect of changing $q_D$ 
at fixed density and $T=0$, and observe behavior similar to
that shown in Fig.~4(d).

In conclusion we have investigated 
a ``glassy ratchet'' effect, which
occurs 
in a glassy system of interacting particles
without a quenched substrate. 
We considered the specific case of a disordered assembly of colloids
with overdamped dynamics where only one
colloid species couples to an ac drive. 
The ac drive is asymmetric, 
with a large positive drive applied for a short time
followed by a small negative drive applied for a longer time, 
and obeys the constraint that the net applied force on the
driven colloids is zero in a single cycle. 
A ratchet effect occurs when the driven and
the non-driven colloids are coupled, producing
a nonlinear velocity-force response.
We find that the dc drift velocity has a maximum value
as a function of the fraction
of driven particles, temperature, and particle density. 
A simple model for the ratchet effect agrees well 
with our simulation results, 
including the occurrence of ratchet reversal regimes.
The system considered here can be realized experimentally 
for magnetic or charged colloids driven
through non-magnetic or uncharged 
colloids. Our results may also have applications for new types 
of electrophoresis devices.

Acknowledgments: We thank E. Weeks,
D.G. Grier, and 
H. Lowen for useful discussions.  
This work was supported by the US Department of Energy under
Contract No. W-7405-ENG-36.

\end{document}